\documentclass[letterpaper]{article} 
\usepackage{aaai24}  

\usepackage{amsmath,amsfonts}
\usepackage{threeparttable}
\usepackage{booktabs}
\usepackage{multirow}
\usepackage{tabularx}
\usepackage{verbatim}

\usepackage{times}  
\usepackage{helvet}  
\usepackage{courier}  
\usepackage[hyphens]{url}  
\usepackage{graphicx} 
\usepackage{natbib}  
\usepackage{caption} 
\usepackage{algorithm}
\usepackage{algorithmic}

\usepackage{newfloat}
\usepackage{listings}
\DeclareCaptionStyle{ruled}{labelfont=normalfont,labelsep=colon,strut=off} 
\floatstyle{ruled}
\newfloat{listing}{tb}{lst}{}
\floatname{listing}{Listing}
\title{Improving PTM Site Prediction by Coupling of Multi-Granularity Structure and
Multi-Scale Sequence Representation}
\author {
    Zhengyi Li\textsuperscript{\rm 1},
    Menglu Li\textsuperscript{\rm 1},
    Lida Zhu\textsuperscript{\rm 1\Large {$_\text{*}$}},
    Wen Zhang\textsuperscript{\rm 1,\rm 2,\rm 3}\thanks{Corresponding authors.}
}

\affiliations {
    \textsuperscript{\rm 1}College of Informatics, Huazhong Agricultural University, Wuhan 430070, China\\
    \textsuperscript{\rm 2}Hubei Key Laboratory of Agricultural Bioinformatics, Huazhong Agricultural University, Wuhan 430070, China\\
    \textsuperscript{\rm 3}Engineering Research Center of Intelligent Technology for Agriculture, Ministry of Education, Wuhan 430070, China\\
    \{lzy\_gnn, mengluli\}@webmail.hzau.edu.cn, \{ldzhu, zhangwen\}@mail.hzau.edu.cn
    
}

\usepackage{bibentry}

\begin{document}

\maketitle

\begin{abstract}
 
Protein post-translational modification (PTM) site prediction is a fundamental task in bioinformatics. Several computational methods have been developed to predict PTM sites. However, existing methods ignore the structure information and merely utilize protein sequences. Furthermore, designing a more fine-grained structure representation learning method is urgently needed as PTM is a biological event that occurs at the atom granularity. In this paper, we propose a \textbf{PTM} site prediction method by \textbf{C}oupling of \textbf{M}ulti-\textbf{G}ranularity structure and \textbf{M}ulti-\textbf{S}cale sequence representation, \textbf{PTM-CMGMS} for brevity. Specifically, multi-granularity structure-aware representation learning is designed to learn neighborhood structure representations at the amino acid, atom, and whole protein granularity from AlphaFold predicted structures, followed by utilizing contrastive learning to optimize the structure representations. Additionally, multi-scale sequence representation learning is used to extract context sequence information, and motif generated by aligning all context sequences of PTM sites assists the prediction. Extensive experiments on three datasets show that PTM-CMGMS outperforms the state-of-the-art methods. Source code can be found at https://github.com/LZY-HZAU/PTM-CMGMS.

\end{abstract}

\section{Introduction}
Post-translational modification (PTM) refers to the biological events of adding small molecular groups to the side chains of amino acids, which greatly enhances the functional diversity of the proteome. Accurate determination of PTM sites contributes to a deepening understanding of protein functions\cite{yang2019succinylation} and their specific roles in various complex cellular processes \cite{Introduction-1,fernando2019s,fang2021histone}, and facilitates precision therapy by illuminating the regulatory systems underlying disease states \cite{cai2022modulating}. Several wet experiments have been used for identifying PTM sites, such as high-sensitivity mass spectrometry, antibodies, isotope labeling, and probes \cite{1}, but they are often time-consuming and labor-intensive, hindering large-scale investigations into PTM.

\begin{figure}[ht]
    \centering
    \includegraphics[width=\columnwidth]{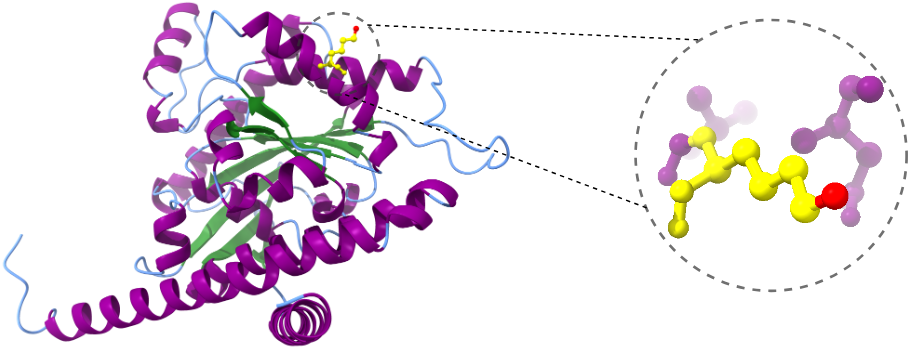}
    \caption{The example of lysine crotonylation, which occurs on the side chain nitrogen atom marked in red. The other atoms of lysine are marked in yellow.}
    \label{environment}
\end{figure}

In recent years, plenty of computational methods have been developed for predicting PTM sites, and these methods are roughly classified into three categories: machine learning-based methods, deep learning-based methods, and pre-trained language model-based methods. The machine learning-based methods \cite{14,2,10,16,6} extracted the protein sequence-derived features manually and then fed them into the classifiers, such as support vector machine (SVM), random forest, and conditional random field algorithm, for PTM site prediction. These methods heavily rely on feature engineering and cannot learn abstract feature representations. Deep learning-based methods \cite{19,18,17,20} employed a deep learning model, such as bidirectional long short-term memory  network(BiLSTM), convolutional neural network(CNN), deep neural network(DNN) and transformer, to learn representations automatically, which have been demonstrated to be powerful techniques for PTM site prediction. However, these methods roughly address PTM site prediction as a sequence modeling problem, ignoring the structure information. The pre-trained language models (PLMs) on large-scale protein sequence corpora have achieved impressive performance for various downstream protein understanding tasks. A series of pre-trained language model-based methods \cite{5,4} have been developed, which utilized the sequence representation extracted from PLMs, such as BERT \cite{40} and ProtTrans \cite{21}, for PTM site prediction. PLMs can learn implicit structure information, but cannot be directly aware of the explicit neighborhood structure of PTM sites. More importantly, PLMs only generate sequence representations at the amino acid or whole protein granularity, which may not be suitable for the specific biological problem as PTM occurs at the atom granularity, shown in Figure~\ref{environment}. 

Recently, researchers have paid more attention to the structure information of PTM sites \cite{22,23}. For example, \citet{24} investigated the structure background of PTM sites and provided detailed annotations on structure features such as secondary structure, tertiary structure context, and solvent accessibility of PTM sites. These studies indicate that PTM sites exhibit structure preferences, which may play a vital role in improving the performance of PTM site prediction. On the other hand, the appearance of AlphaFold \cite{25, 26} enables us to obtain protein structures of a wide range of organisms without resorting to traditional experimental techniques, inspiring us to design a structure-aware PTM site prediction model.

In this paper, we propose a novel PTM site prediction method by Coupling of Multi-Granularity Structure and Multi-Scale Sequence Representation, namely PTM-CMGMS. Specifically, we design a multi-granularity structure-aware representation learning, including the atom, amino acid, and whole protein granularity, to learn neighborhood structure representations of PTM sites from AlphaFold predicted structures. Further, we adopt contrastive learning to optimize the structure representations. Additionally, we use a multi-scale sequence representation learning to extract context sequence information of PTM sites, and obtain the motif-based information by detecting the differences between context sequences and motif which is generated by aligning all context sequences of PTM sites. Finally, the combination of structure representations, context sequence information, and motif-based information is fed into an MLP to produce the prediction probability.

In summary, the main contributions of this paper are described as follows:

\begin{itemize}
    \item We design a multi-granularity structure-aware representation learning that integrates structure information of amino acid, atom, and whole protein granularity to learn more comprehensive structure representations. To the best of our knowledge, this is the first method to model PTM sites at the atom granularity, which is more consistent with biochemical facts.
   
    \item We utilize contrastive learning to optimize structure representations of PTM sites for enhancing the generalization capacity of PTM-CMGMS.

    \item We design a multi-scale sequence representation learning to simulate local amino acid interactions at different scales for obtaining sequence representations, which is used to assist the prediction of our model.
\end{itemize}

\section{Related Work}

Existing methods mainly focus on the context sequence information of PTM sites.

A line of works manually extract the sequence information as feature vectors and train a classic classifier to predict PTM sites \cite{13,15,11,12,9,7,3,8}. For example, \citet{2} design an encoding scheme called Position Weighted Amino Acid Composition (PWAA) and use an SVM as the classifier for PTM site prediction. PSuccE \cite{10} adopts an ensemble learning algorithm with a feature selection scheme to enhance the prediction performance. LightGBM-CroSite \cite{6} integrates multiple encoding schemes to extract more comprehensive sequence features and uses the elastic net to remove redundant information. Although machine learning-based methods have achieved great success in PTM site prediction, they heavily rely on hand-crafted features and domain expertise, making it difficult to capture the implicit information of protein sequences. 
The other line of works usually train a deep learning model that learn representations automatically. For instance, DeepNitro \cite{19} constructs four types of encoding features (positional amino acid distributions, sequence contextual dependencies, physicochemical properties, and position-specific scoring features), and uses a DNN to learn the high-dive information
for representing the PTM sites. Deep-Kcr \cite{17} combines sequence-based features, physicochemical property-based features, and numerical space-derived information to a CNN for further representation learning to predict PTM sites. Similar to Deep-Kcr, DeepSuccinylSite \cite{18} adopts one-hot and word embeddings as features and inputs them into a CNN for PTM site prediction. 
Adapt-Kcr \cite{20} utilizes an adaptive word embedding algorithm to encode the sequence, employs CNN, a BLSTM, and an attention mechanism to better capture the latent information of adaptive embedding for predicting PTM sites. These deep learning-based methods merely utilize the protein sequence while ignoring structure information.

Additionally, there are also a series of works based on pre-trained language models. BERT-Kcr \cite{5} transfers each amino acid into a word as the input of BERT \cite{40}, and adopts BiLSTM as the classifier for PTM site prediction. They also attempted to pre-train BERT using protein sequences and fine-tune the model for this prediction task. LMSuccSite \cite{4} utilizes the embeddings derived from ProtTrans in conjunction with supervised word embedding to improve the prediction performance. The pre-trained language models typically involve tens of millions of parameters, and require too much time and computational resources, especially for large-scale sequences input. Additionally, the pre-trained language models can not generate representations of atom granularity for PTM site prediction.

\begin{figure*}[ht]
    \centering
    \includegraphics[width=\textwidth]{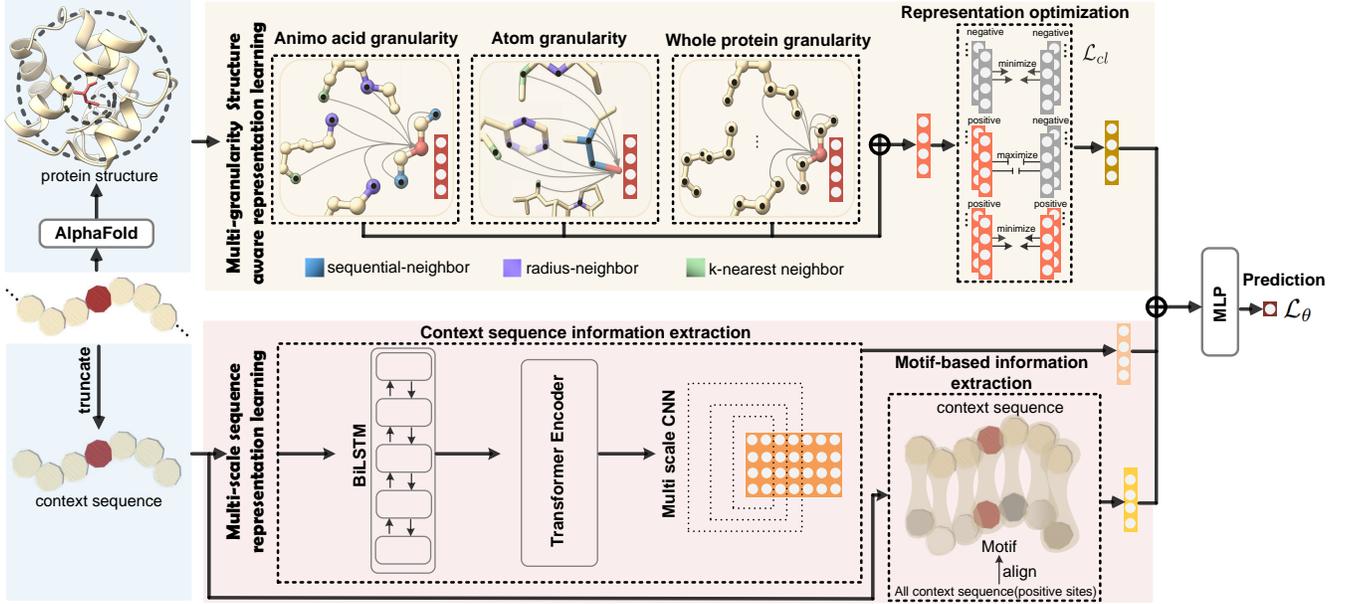}
     \caption{The overview of our proposed PTM-CMGMS.}
     \label{Framwork}
\end{figure*}

Different from the above methods, we integrate multi-granularity structure and multi-scale sequence information to construct a PTM site prediction method. 

\section{Method}
	In this section, we first formulate the PTM site prediction problem. After that, we elaborately enumerate all components of the proposed method PTM-CMGMS, shown in Figure~\ref{Framwork}. At last, we provide an exposition of model training.
    \subsection{Problem Formulation}
    
    Given a protein sequence $S=\{s_i|i=1,2,..., L\}$ with $L$ residues. $\mathcal{X}=\{x_i\in\mathbb{R}^{{Z}\times 3}|i=1,2,...,L\}$ denotes the coordinates in 3D space for the residues with $Z$ atoms (e.g., $N$, $C_{\alpha}$, $C$, $O$) obtained by AlphaFold. The context sequence $P$ is obtained by truncating 15aa short peptide fragments upstream and downstream of $s_i$ in $S$, where the residue $s_i$ is to be predicted, and the length of $P$ is 31aa. In this paper, our goal is to construct a model that predicts whether $s_i$ is a PTM site (positive) or not (negative) based on the protein structure and context sequence, which is a binary classification problem.

    \subsection{Multi-granularity Structure-aware Representation Learning}
    In this section, we learn multi-granularity structure representations of $s_i$ at the amino acid, atom, and whole protein granularity. Then contrastive learning is utilized to optimize the structure representations.
    \subsubsection{Amino acid granularity.}

    We capture three types of neighbors within residue $s_i$ to learn the neighborhood structure representation at amino acid granularity, including sequential-neighbors, radius-neighbors, and k-nearest neighbors \cite{27,28}. They are generated as follows: sequential-neighbors $\{s_j|\ |s_j-s_i|<d_{seq}\}$, where $|s_j-s_i|$ represents the sequential distance between residues $s_j$ and $s_i$ at $S$; radius-neighbors $\{s_j|D(x_j(C_\alpha)-x_i(C_\alpha))< d_{spa}\}$, where $D(x_j(C_\alpha)-x_i(C_\alpha))$ represents the Euclidean distance between $C_\alpha$ of residues $s_j$ and $s_i$; $k$-nearest neighbors $\{s_j\}_{j=1}^k$, where $s_j$ is the top $j$ nearest neighbors of $s_i$ based on the Euclidean distance. In this study, we set $d_{seq}=3$, $d_{rad}=10\mathring{A}$ and $k=10$ \cite{28}. 
    
    Based on these neighbors, we can obtain the amino acid granularity representation of the residue $s_i$:

     \begin{equation}
    	r_{s_i}^{aa}=\sigma(W\sum_{j\in \mathcal{N}(s_i)}(H_j||E_j))
    \end{equation}
    where $W$ is the learnable parameter, $\sigma(\cdot)$ denotes an activation function LeakyReLU, $||$ represents concatenation operation, and $\mathcal{N}(s_i)$ is a set of three type neighbors of residue $s_i$. $H_j$ represents the embedding of residue type, secondary structure, backbone torsion angle $\phi$ and $\psi$, and relative solvent accessibility of residue $s_j$ \cite{29}. $E_j$ represents the embedding of specific neighbor type of $s_j$ related to $s_i$, the sequential and Euclidean distance between $s_j$ and $s_i$.

    \subsubsection{Atom granularity.} We capture three types of neighbors within the particular side chain atom of residue $s_i$ to learn the neighborhood structure representation at atom granularity, which is due to PTM occurring at the particular side chain atom of residue $s_i$ (e.g. crotonylation occurs on the side chain nitrogen atom). Here, we take the prediction of whether $s_i$ is a crotonylation site as an example to illustrate. For the residue $s_i$, the generation process of three types of neighbors of side chain nitrogen atom is similar to the amino acid granularity, the atom granularity representation is as follows:
    
    \begin{equation}
    	 r_{s_i}^{atom}=\sigma(W\sum_{j\in \mathcal{N} ({s_i}^{(N)})}(h_j||e_j))
    \end{equation}
    where $h_j$ represents the embedding of atom type of $j$ and its residue type, $e_j$ represents the embedding of specific neighbor type of $j$ related to the nitrogen atom of $s_i$, the sequential and Euclidean distance between them. $s_i^{(N)}$ represents the nitrogen atom of $s_i$.
     
    \subsubsection{Whole protein granularity.} We capture all residue information to learn the neighborhood structure representation at whole protein granularity, and the whole protein granularity representation of $s_i$ is as follows: 
    \begin{equation}
    	r_{s_i}^{pro}=\sigma(W\sum_{j\in S}(H_j))
    \end{equation}
    
    After that, we integrate these three representations to obtain the multi-granularity structure representation of $s_i$, denoted as $r^{MG}$:
    
     \begin{equation}
    	r^{MG}=\text{MLP}(r_{s_i}^{atom}||r_{s_i}^{aa}||r_{s_i}^{pro})
    \end{equation}
    where MLP is a multi-layer perceptron.
  
    \subsubsection{Representation optimization.} 
    In order to learn discriminative multi-granularity structure representations, we introduce contrastive learning in our model. For the structure representations of residue $s_i$ and $s_j$, we calculate the contrastive learning loss as follows \cite{30}: 
    \begin{equation}
    	\begin{aligned}
    			\mathcal{L}_{cl}=\frac{1}{2}(&(1-Y)D(r_{s_i}^{MG},r_{s_j}^{MG})^2+\\
    			&Y\{\max{(0,M-D(r_{s_i}^{MG},r_{s_j}^{MG}))^2}\})
    	\end{aligned}
    \end{equation}
    where $Y$ is equal to 0 if and only if $s_i$ and $s_j$ belong to the same label (both positive or both negative samples), and 1 in other cases. $M$ represents the cut-off distance. If $s_i$ and $s_j$ belong to different labels and the distance of their structure representations is less than $M$, we optimize them so that they are far away from each other.

    \subsection{Multi-scale Sequence Representation Learning}
    In this section, we learn multi-scale sequence representation
    by integrating context sequence information and motif-based information. 
    \subsubsection{Context sequence information extraction.}

    For the context sequence $P$ of $s_i$, we initialize embedding vectors for each residue by fusing position and token embeddings \cite{31}. The embedding of $P$ is computed as follows:
    \begin{equation}
     r^P=embedding_{(position)}+embedding_{(token)}
    \end{equation}
    where $r^P\in\mathbb{R}^{(31\times d)}$, $d$ represents the embedding dimension.
    
    We then utilize BiLSTM \cite{32} to integrate forward and reverse context sequence information, and feed the output of BiLSTM into the transformer encoder \cite{33} for further feature extraction. The core of the transformer encoder is the multi-head self-attention mechanism, which enables $s_i$ to aggregate other residue embeddings with different attention weights. 

    \begin{equation}
      	Q_i=r^PW_i^Q;K_i=r^PW_i^K;V_i=r^PW_i^V
    \end{equation}

    \begin{equation}
      	\begin{aligned}
      	\text{head}_i=\text{Softmax}(\frac{Q_iK_i^T}{\sqrt{d}})V_i
      	\end{aligned}
    \end{equation}
     
    \begin{equation}
     	r^P=\|_{i=1}^h\text{head}_i\cdot W
    \end{equation}
    where $W_i^Q$, $W_i^K$, $W_i^V$ are query, key, value matrix of $i$-th head ($i=1,2,...,h$), respectively, $h$ is the number of heads, $W$ is a linear transformation matrix, and $d$ is the scaling factor. 

    In order to capture local patterns, we further design a multi-scale CNN (MCNN) with multiple convolutional kernels $\{c_i\}_{i=1}^C$ to simulate the local interaction of residues at different scales.

     \begin{equation}
     	r^P=\sigma(W(\mathrm{mean}(\|_{i=1}^C\text{Pool}(\text{Conv}(r^P,c_i)))))
     \end{equation}
    
    Based on the combination of BiLSTM, Transformer encoder, and MCNN, we extract the context sequence information to obtain the representations $r^P$ of context sequence $P$.

    \subsubsection{Motif-based information extraction.}
    We align the context sequences of all known positive sites in the training set to obtain the frequency of residues at each position. The residues with the highest frequency (denoted as $w$) at each position make up the motif $M$ (the length is 31aa) which reflects the contextual pattern of the PTM sites. In general, $s_i$ is more likely to belong to a PTM site if context sequence $P$ is similar to $M$. 
    
    Here, we adopt physicochemical properties AAIndex (AAI) and evolutionary relationships BLOSUM62 \cite{34} to characterize the difference between $P$ and $M$: 

    \begin{equation}
    \begin{aligned}
        r^M=\text{MLP}(w\cdot D( & \text{AAI}_M||\text{BLOSUM62}_M,\\
      		& \text{AAI}_P||\text{BLOSUM62}_P))
    \end{aligned}
	\end{equation}
    where $\text{AAI}_M$ and $\text{AAI}_P$ denote the physicochemical property embedding matrices of $M$ and $P$, $\text{BLOSUM62}_M$ and $\text{BLOSUM62}_P$ denote the evolutionary relationship embedding matrices of $M$ and $P$, respectively. 
    
    After that, we combine these two representations to obtain a multi-scale sequence representation of $P$:	
    \begin{equation}
		r^{MS}=r^P || r^M
	\end{equation}

    \subsection{Model Training}
    For the residue $s_i$, the multi-granularity structure representations $r^{MG}$ and multi-scale sequence representation $r^{MS}$ are concatenated and fed into an MLP to get the probability of PTM site:

    \begin{equation}
    	\hat{y}=\text{MLP}\left(r^{MG}||r^{MS}\right)
    \end{equation}
    
    The training objective is to minimize the loss function:
 	
 	\begin{equation}
 		\mathcal{L}_\theta=-\sum_{i=1}^N\left(y_i\log\left(\hat{y}_i\right)+\left(1-y_i\right)log\left(1-\hat{y}_i\right)\right)
 	\end{equation}
	where $N$ represents the total number of samples in the training set, 
    $y_i$ represents the true label of the $i$-th sample, and $\hat{y}_i$ represents the probability predicted by our model. 
 
\section{Experiments}
	In this section, we first introduce the experimental settings, and then compare our proposed PTM-CMGMS with baselines. We then employ the ablation study to investigate the effectiveness of each component. Further, we also explore the importance of structure information for PTM site prediction.
 
    \subsection{Experimental Settings}
    \subsubsection{Datasets.}
    To test the performance of our proposed PTM-CMGMS, we evaluate it on three publicly available datasets with different PTM types, including Crotonylation \cite{5,35}, Succinylation \cite{4}, and Nitrosylation \cite{36}. The division scheme of these datasets is consistent with the previous studies, and the redundant sites in the datasets are removed. The statistics of three datasets are in Table~\ref{datasets}.
    
    \begin{table}[h]
    \centering
    \begin{tabular*}{\columnwidth}    
        {@{\extracolsep{\fill}}ccccc@{\extracolsep{\fill}}}
        \toprule
        \multirow{2}{*}{Datasets} & \multicolumn{2}{c}{Training set} & \multicolumn{2}{c}{Test set} \\
        & Positive & Negative & Positive & Negative \\
        \midrule
        Crotonylation & 6,975 & 6,975 & 2,889 & 1,939 \\
        Succinylation & 4,749 & 4,750 & 253 & 2,973 \\
        Nitrosylation & 3,276 & 3,276 & 351 & 3,168 \\
        \bottomrule 
    \end{tabular*}
    \caption{The statistics of three datasets.}%
    \label{datasets}
    \end{table}

    \subsubsection{Baselines.}
    We compare our proposed PTM-CMGMS with several baselines, which can be categorized as follows.
    \begin{itemize}
    \item PSuccE \cite{10} adopts an ensemble learning algorithm to predict succinylation sites combining multiple features with a feature selection scheme (information gain). 
    \item LightGBM-CroSite \cite{6} uses the elastic net to select the optimal feature subset from multiple sequence features, and inputs them into a LightGBM to predict crotonylation sites.
    \item DeepNitro \cite{19} utilizes four types of encoding features, and trains a deep neural network to predict nitrosylation sites. 
    \item Deep-Kcr \cite{17} combines sequence-based, physicochemical property-based, and numerical space-derived information to predict crotonylation sites based on a CNN.
    \item DeepSuccinylSite \cite{18} combines one-hot and word embeddings as features, and uses a CNN to predict succinylation sites.
    \item Adapt-Kcr \cite{20} utilizes adaptive embedding and is based on a CNN together with a BiLSTM and attention architecture to predict crotonylation sites.
    \item BERT-Kcr \cite{5} employs pre-trained BERT to transfer each amino acid into a word, and adopts a BiLSTM to predict crotonylation sites. 
    \item LMSuccSite \cite{4} combines supervised word embedding and embedding learned from a pre-trained protein language model, and uses a neural network to predict succinylation sites.
    \end{itemize}
         
    \subsubsection{Implementation details.}
    To measure the performance of our method, we implement the 10-fold cross-validation on the training set, conduct independent testing on the test set, in which the model is trained on the training set. And we adopt three evaluation metrics, including the Matthews correlation coefficient (MCC), the area under the receiver-operating characteristic curve (AUC), and the area under the precision-recall curve (AUPR). The experiments are performed on the machine with Intel(R)Core(TM)i9-10980XE CPU@3.00GHz and 2 GPUs(NVIDIA GeForce RTX 3090). 
       
    \begin{table*}[h]
    \centering
    \begin{tabular*}{\textwidth}    
        {@{\extracolsep{\fill}}lccccccccc@{\extracolsep{\fill}}}
        \toprule
        \multirow{2}{*}{Method} & \multicolumn{3}{c}{Crotonylation} & \multicolumn{3}{c}{Succinylation} & \multicolumn{3}{c}{Nitrosylation} \\
        &	MCC & AUC &	AUPR &	MCC & AUC &	AUPR &	MCC & AUC &	AUPR \\
        \midrule
        PSuccE & 0.5142 & 0.8427 & 0.8744 & 0.1665 & 0.7187 & 0.1839 & 0.2699 & 0.7807 & 0.2509 \\
        LightGBM-CrotSite & 0.5396 & 0.8587 & 0.8858 & 0.1803 & 0.7372 & 0.1887 & 0.2562 & 0.7687 & 0.2221 \\
        \midrule
        DeepNitro & 0.5982 & 0.8873 & 0.9082 & 0.2831 & 0.8073 & 0.2837 & 0.2344 & 0.7544 & 0.2193 \\
        Deep-Kcr & 0.5004 & 0.8445 & 0.8734 & 0.2009 & 0.7307 & 0.1944 & 0.2575 & 0.7625 & 0.2257 \\
        DeepSuccinylSite & 0.6422 & 0.9011 & 0.9181 & 0.2272 & 0.7753 & 0.2136 & 0.1629 & 0.6845 & 0.1790 \\
        Adapt-Kcr & \underline{0.6426} & \underline{0.9020} & \underline{0.9188} & 0.2360 & 0.7782 & 0.2258 & 0.2064 & 0.7311 & 0.1999 \\
        \midrule
        BERT-Kcr & 0.5649 & 0.8660 & 0.8942 & 0.2123 & 0.7575 & 0.2190 & 0.2149 & 0.7489 & 0.2284 \\
        LMSuccSite & 0.6222 & 0.8914 & 0.9078 & \underline{0.2990} & \underline{0.8169} & \underline{0.2966} & \underline{0.3060} & \underline{0.8112} & \underline{0.2850} \\
        \midrule
        PTM-CMGMS & \textbf{0.7005} & \textbf{0.9245} & \textbf{0.9386} & \textbf{0.3072} & \textbf{0.8306} & \textbf{0.3024} & \textbf{0.3398} & \textbf{0.8369} & \textbf{0.3136} \\
        \bottomrule 
    \end{tabular*}
    \caption{Comparison results of PTM-CMGMS and baselines on three datasets (Crotonylation, Succinylation, and Nitrosylation). Note that the highest score in each column is in bold and the second-best score is underlined. }%
    \label{compare}
    \end{table*}
    
    \begin{figure*}[h]
    \centering
    \includegraphics[width=\textwidth]{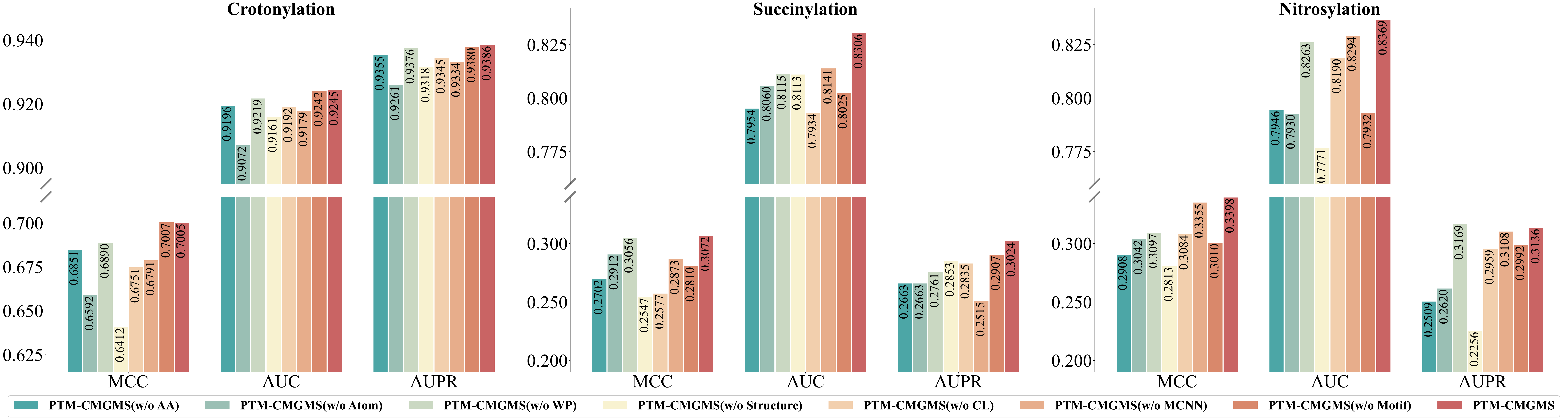}
     \caption{Results of PTM-CMGMS and its variants on three datasets.}
     \label{ablation}
    \end{figure*}
    
    \subsection{Comparison with Baselines}
    Table~\ref{compare} shows the evaluation results of PTM-CMGMS and the baselines on three datasets. According to the results shown in Table~\ref{compare}, PTM-CMGMS achieves the best performance on three datasets. We also have the following observation: (1) Compared with PSuccE and LightGBM-CrotSite which rely on hand-crafted features, PTM-CMGMS significantly exceeds all these baselines on three datasets, deep learning-based methods (DeepNitro, Deep-Kcr, DeepSuccinylSite, and Adapt-Kcr) and pre-trained language model-based methods (BERT-Kcr and LMSuccSite) also achieve better performance than PSuccE and LightGBM-CrotSite on most datasets, which indicates that using deep learning techniques to automatically extract sequence information can enhance the prediction for PTM sites. (2) Compared with DeepNitro, Deep-Kcr, DeepSuccinylSite, and Adapt-Kcr which only consider protein sequence information, PTM-CMGMS achieves the best performance on all datasets, which implys that considering structure information benefits PTM site prediction. (3) Compared with BERT-Kcr and LMSuccSite that utilize representations at the amino acid granularity generated by pre-trained language models, PTM-CMGMS makes improvements of 13.56\% and 7.83\% on Crotonylation dataset, 9.49\% and 0.82\% on Succinylation dataset, and 12.49\% and 3.38\% on Nitrosylation dataset in terms of MCC, indicating the information of atom granularity is also beneficial for PTM sites prediction.
   
    \subsection{Ablation Study}
    To illustrate the importance and effectiveness of each component designed in our model, we consider the following variants of PTM-CMGMS.
    \begin{itemize}
        \item PTM-CMGMS without amino acid granularity (w/o AA) removes the structure representations at the amino acid granularity in the multi-granularity structure-aware representation learning.
        \item PTM-CMGMS without atom granularity (w/o Atom) removes the structure representations at the atom granularity in the multi-granularity structure-aware representation learning.
        \item PTM-CMGMS without whole protein granularity (w/o WP) removes the structure representations at the whole protein granularity in the multi-granularity structure-aware representation learning.
        \item PTM-CMGMS without contrastive learning (w/o CL) removes representation optimization based on contrastive learning in the multi-granularity structure-aware representation learning.
        \item PTM-CMGMS without structure information (w/o Structure) removes multi-granularity structure-aware representation learning.
        \item PTM-CMGMS without multi-scale CNN (w/o MCNN) removes the multi-scale CNN in multi-scale sequence representation learning.
        \item PTM-CMGMS without Motif-based information extraction (w/o Motif) removes the motif-based information extraction in multi-scale sequence representation learning.
    \end{itemize}

    As shown in Figure~\ref{ablation}, PTM-CMGMS with the integration of multi-granularity structure and multi-scale sequence representation learning achieves superior performances, and the removal of any will undermine the predictive capacity of PTM-CMGMS on most datasets. Besides, we have the following observations: (1) PTM-CMGMS outperforms the PTM-CMGMS(w/o Structure) and PTM-CMGMS(w/o MCNN), which demonstrates that the combination of multi-granularity structure-aware information and multi-scale sequence information helps improve the performance of PTM site prediction. In addition, PTM-CMGMS(w/o Structure) gets an obvious performance drop, indicating the prediction performance is boosted mostly by the structure information and multi-granularity structure-aware representation learning is the core component of the model architecture. (2) PTM-CMGMS outperforms the PTM-CMGMS(w/o AA), and PTM-CMGMS(w/o WP), which demonstrates that structure representations at these granularities are all helpful for PTM site prediction. Note that PTM-CMGMS outperforms the PTM-CMGMS without the information of atom granularity (PTM-CMGMS(w/o Atom)), which verifies the rationality to model the PTM site into finer-grained atom level. These granularity information provide a more comprehensive characterization of the structure feature of PTM sites. (3) PTM-CMGMS(w/o CL) also gets a performance drop on all metrics, which shows that the discriminative structure representations obtained by contrastive learning can provide useful information for PTM site prediction. (4) PTM-CMGMS performs better than PTM-CMGMS(w/o Motif), which implies that the motif generated by aligning all context sequences of PTM sites can assist the prediction. 

\begin{figure}[h]
    \centering
    \includegraphics[width=\columnwidth]{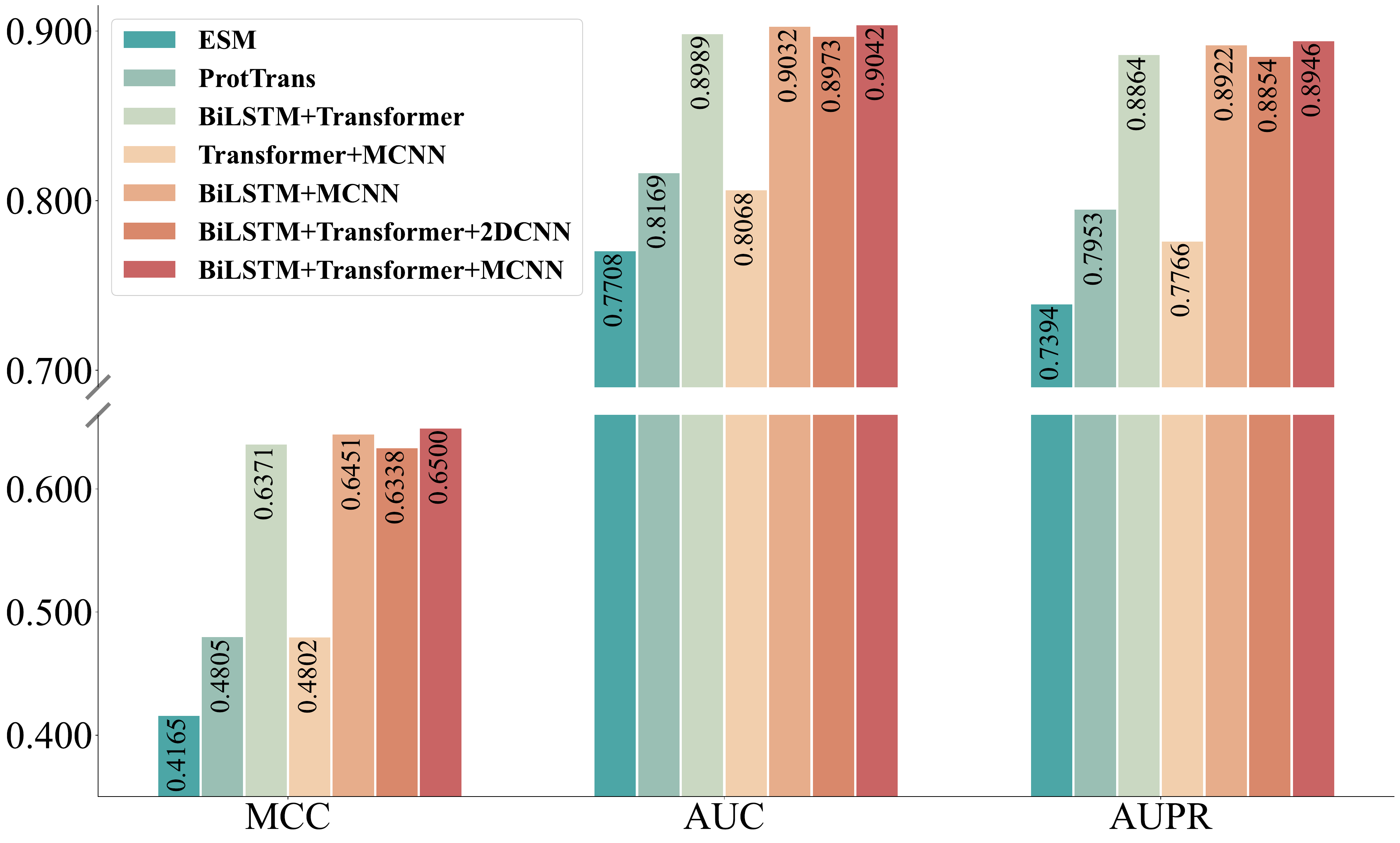}
     \caption{Results of various context sequence information extraction architectures on the Crotonylation dataset.}
     \label{seq_analysis}
    \end{figure}

    To explore the effectiveness of the combination of BiLSTM, Transformer, and MCNN (BiLSTM+Transformer+MCNN), we also investigate the performance of various combination of context sequence information extraction architectures on the Crotonylation dataset, including BiLSTM+Transformer, Transformer+MCNN, BiLSTM+MCNN, and BiLSTM+Transformer+2DCNN that replaces the MCNN with a 2DCNN. The PLMs pre-trained on large-scale protein sequence corpora have achieved impressive performance on various downstream protein understanding tasks, thus ESM\cite{lin2022language} and ProtTrans are also within our consideration. Results in Figure~\ref{seq_analysis} show that BiLSTM+Transformer+MCNN achieves the best performance, which indicates that the combination of BiLSTM, Transformer, and MCNN can comprehensively capture sequence information to promote PTM site prediction. Besides, we have the following observation: (1) BiLSTM+Transformer+MCNN is superior to PLMs (ESM and ProtTrans), which demonstrates the effectiveness of our designed multi-scale sequence information extraction strategy on PTM site prediction. (2) BiLSTM+Transformer+MCNN outperforms any combination of pairwise architectures, this may be due to the complementary strengths of the three architectures. (3) BiLSTM+Transformer+2DCNN that replaces the MCNN with a 2DCNN gets a performance drop on all metrics, suggesting that MCNN may be more suitable for extracting local information of context sequences for PTM site prediction. 
    
      \subsection{Effectiveness of structure information for PTM site prediction } 
    
    To verify the effectiveness of structure information for predicting PTM sites, we compare the performance of PTM-CMGMS and Adapt-Kcr (a baseline that achieves the best performance on the Crotonylation dataset) in dealing with PTM sites with different numbers of non-local contact residues. Residues are considered as non-local contact with the PTM sites if they are separated with greater than 20 residues in the primary structure and their spatial distance is less than 12\AA\ \cite{37,38}. Structure information of PTM sites mainly refers to these non-local contact residues. We have the following conclusions: (1) PTM-CMGMS consistently surpasses Adapt-Kcr when the number of non-local contact residues is greater than 0, which suggests that using of structure information can effectively enhance the prediction for PTM sites with multiple non-local contact residues. (2) PTM-CMGMS is also better than Adapt-Kcr when the number of non-local contact residues is equal to 0 (i.e. the structure information of PTM sites is relatively insufficient), implying the advantage of integrating both aspects of structure and sequence information. These results demonstrate the effectiveness of structure information for PTM site prediction.

    \begin{figure}[h]
        \centering
        \includegraphics[width=\columnwidth]{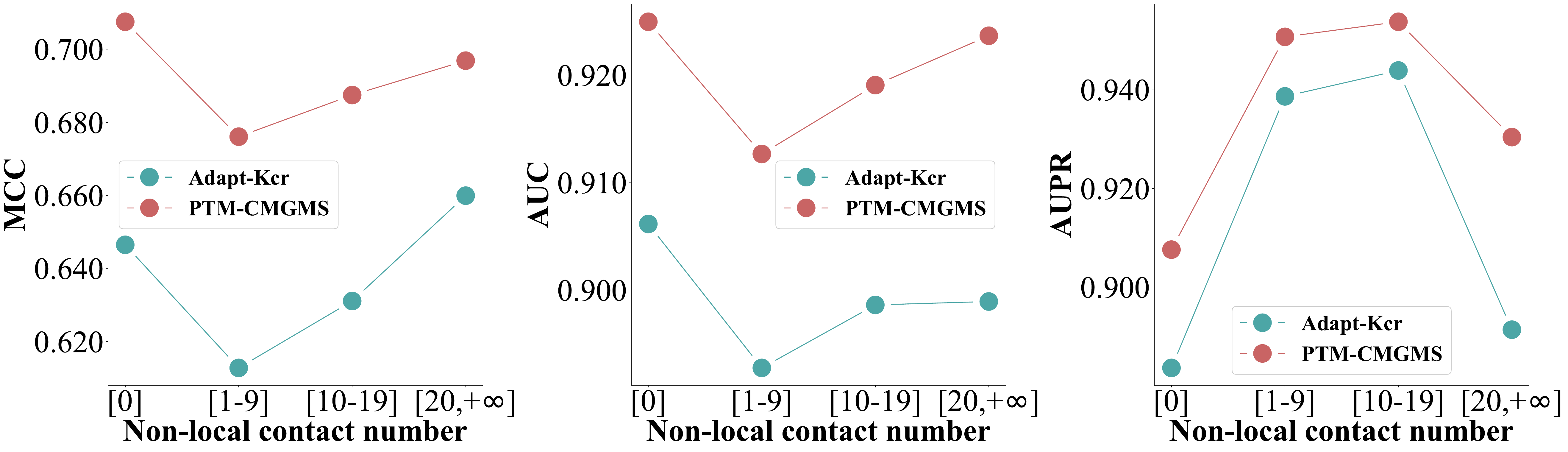}
        \caption{Results of PTM-CMGMS and Adapt-Kcr in dealing with PTM sites with different numbers of non-local contact residues on the Crotonylation dataset.}
        \label{contacts}
    \end{figure}

\section{Conclusion}
	In this paper, we propose a PTM site prediction method (PTM-CMGMS), which couples the multi-granularity structure and multi-scale sequence representation. PTM-CMGMS designs a multi-granularity structure-aware representation learning(including the atom,
amino acid, and whole protein granularity) to capture neighborhood structure representations of PTM sites based on AlphaFold predicted structures, and adopts contrastive learning to optimize structure representations. Further, PTM-CMGMS leverages a multi-scale sequence representation learning to extract context sequence information of PTM sites, and utilizes the motif generated by aligning all context sequences of positive sites to assist the prediction. Experimental results demonstrate the effectiveness of PTM-CMGMS and the importance of multi-granularity structure and multi-scale sequence information for PTM site prediction.

	In future work, we have several directions to optimize the PTM site prediction method, such as considering crosstalk between different types of PTMs and providing interpretable PTM site prediction method.

\section{Acknowledgments}
This work was supported by the National Natural Science Foundation of China (62372204, 62072206, 61772381,
62102158); Huazhong Agricultural University Scientific \& Technological Self-innovation Foundation; Fundamental Research Funds for the Central Universities
(2662021JC008, 2662022JC004). The funders have no role
in study design, data collection, data analysis, data interpretation or writing of the manuscript.

\bibliography{aaai24}

\end{document}